\documentclass[showpacs,prl,twocolumn,aps]{revtex4}

\usepackage[T1]{fontenc}
\usepackage[latin1]{inputenc}
\usepackage{bm}
\usepackage{amsmath}
\usepackage{amssymb}
\usepackage{latexsym}
\usepackage{amsfonts}
\usepackage{epsfig}
\usepackage{psfrag}

\newcommand{\f}[1]{\mbox{\boldmath$#1$}}

\newcommand{\na}{\mbox{\boldmath$\nabla$}}
\newcommand{\bea}{\begin{eqnarray}}
\newcommand{\ea}{\end{eqnarray}}
\newcommand{\eea}{\end{eqnarray}}

\begin{document}

\title{Vortex quantum creation and winding number scaling in a
  quenched spinor Bose gas} 

\author{Michael Uhlmann$^1$, Ralf Sch\"utzhold$^{1,*}$ and Uwe
R.~Fischer$^{2,\dagger}$}

\affiliation{$^1$Institut f\"ur Theoretische Physik,
Technische Universit\"at Dresden, D-01062 Dresden, Germany
\\
$^2$Eberhard-Karls-Universit\"at T\"ubingen,
Institut f\"ur Theoretische Physik\\
Auf der Morgenstelle 14, D-72076 T\"ubingen, Germany}

\begin{abstract}
Motivated by a recent experiment, we study non-equilibrium quantum
phenomena taking place in the quench of a spinor Bose-Einstein
condensate through the zero-temperature phase transition separating
the polar paramagnetic and planar ferromagnetic phases.
We derive the typical spin domain structure
(correlations of the effective magnetization) created by the quench
arising due to spin-mode quantum fluctuations, and establish a
sample-size scaling law for the creation of spin vortices, which are
topological defects in the transverse magnetization.
\end{abstract}


\maketitle

A rapid sweep through a symmetry-breaking zero-temperature phase
transition entails fascinating non-equilibrium quantum phenomena. 
The quench generates causally disconnected spatial domains of
different order parameter values (corresponding to 
positions in the new ground-state manifold), 
which may ultimately form topological defects in a quantum
version of the Kibble-Zurek mechanism \cite{KZ,Hindmarsh,ZDZ}.
Apart from its relevance in condensed-matter theory 
(e.g., in superconductors, these defects affect measurable
transport properties like conductivity and susceptibility)
and potentially cosmology \cite{Hindmarsh}, this phenomenon is very
interesting from a fundamental point of view, since the defect
distribution directly originates from the initial quantum fluctuations
and thus maps out their properties (such as the ground-state entanglement). 

Because of their comparably long response times, dilute atomic gases
provide a unique opportunity for exploring these fundamental quantum
many-body phenomena far from equilibrium in the laboratory. 
In the following, we derive the spectrum of fluctuations induced by a 
rapid quench to the (planar) ferromagnetic state of a spinor Bose gas, 
and establish a general sample-size dependent scaling law for the
resulting variance of the net number of spin vortices 
(i.e., of the winding number).  
Besides the total defect density, the winding number within a given
area determines the spectrum of the created magnetization
fluctuations. 
Thus dilute Bose gases serve as a useful and experimentally accessible toy
model for the general case and may allow us to extract universal
properties of such dynamical phase transitions. 

Spinor Bose-Einstein condensates are created by trapping different
hyperfine states of a particular atomic species by optical means
\cite{Stenger}, which enables, {\em inter alia}, the investigation of 
coherent spin-exchange dynamics \cite{M-S_Chang}, and the formation
of spin domains by a dynamical instability \cite{Zhang}.
Non-equilibrium phenomena in a spinor Bose gas were realized in a
recent experiment \cite{Sadler}, where an initially paramagnetic state
was rapidly quenched through a quantum phase transition to a final
ferromagnetic state \cite{polar}. 
Such a quench results in spin vortices,  
(imaged {\em in situ} by phase contrast techniques \cite{Higbie}), 
which are topological defects in the magnetization with a paramagnetic
core \cite{Saito,Chiral}.  
The dilute spinor Bose gas can be described in terms of multi-component
field operators $\hat\psi_a$, whose dynamics is governed
by the Hamiltonian density ($\hbar=1$) 
\bea
\hat{\cal H} & = &
\frac1{2m}(\na\hat\psi_a^\dagger)\cdot\na\hat \psi_a
+V_{\rm trap}\hat\psi^\dagger_a\hat \psi_a
+\frac{c_0}2 \hat\psi^\dagger_a \hat\psi^\dagger_b
\hat\psi_b \hat \psi_a
\nonumber\\
& & +\frac{c_2}2 \f{F}_{ab}\cdot\f{F}_{cd}\, \hat \psi^\dagger_a  
\hat \psi^\dagger_c \hat\psi_b \hat\psi_d
-q \hat \psi^\dagger_{z} \hat\psi_{z}\,,
\label{Hamilton}
\ea
where $m$ denotes the mass of the atoms.
The vector $\f{F}_{ab}$ contains the spin matrices \cite{Ohmi} and
determines the effective magnetization 
$\hat{\f{F}}=\f{F}_{ab}\hat \psi^\dagger_a\hat\psi_b/\varrho$, where 
a summation over component indices $a,b,c,d$ is implied and
$\varrho = \langle \hat\psi^\dagger_a \psi_a\rangle$
is the total (conserved) density. 
For spin-one systems, the sum runs over $a=0,\pm1$ or, alternatively,
over $a=x,y,z$ with $\psi_0 = \psi_z$ and
$\psi_\pm=(\psi_x\pm i\psi_y)/\sqrt{2}$.
The coupling constants
$c_0= 4\pi (a_0 + 2a_2)/(3m)$ and $c_2 = 4\pi (a_2-a_0)/(3m)$
are determined by scattering lengths $a_S$ for the scattering channel
with total spin $S$.
For $c_0\gg|c_2|$, density fluctuations are energetically suppressed
in comparison with the spin modes.
Since we are interested in the effects of the phase transition 
(which is, strictly speaking, only well-defined for infinite systems) 
and not in the impact of inhomogeneities, we assume
$\varrho \approx \rm const.$,\ and hence omit the scalar trapping potential
$V_{\rm trap}$. 
Finally, $q$ denotes the strength of the quadratic 
(one-particle) Zeeman shift $qF_z^2$ 
\cite{Sadler}, where an external magnetic
field is oriented along the $z$ direction.
(The linear Zeeman shift can be eliminated by transforming to the
co-rotating frame, because the Larmor precession is much faster 
than all other frequency scales \cite{Saito}.) 

Assuming $c_2<0$, the system is ferromagnetic for vanishing Zeeman
shift $q=0$, i.e., the magnetization $\hat{\f{F}}$ assumes a maximum
value $\langle\hat{\f{F}}\rangle^2=1$ in some given but arbitrary
direction (broken-symmetry phase).
For small but non-zero $q$, the magnetization
$\langle\hat{\f{F}}\rangle$ lies in the $x,y$-plane due to the term
$qF_z^2$ but is still maximal, $\langle\hat{\f{F}}\rangle^2=1$.
For large $q$, however, the ground state corresponds to a condensate
in the $\psi_0 = \psi_z$ component only and symmetry is restored,
i.e., the magnetization $\langle\hat{\f{F}}\rangle$ vanishes
(``paramagnetic'' phase). 
Hence, by rapidly lowering $q$, we may quench the system from an initially
paramagnetic to an (effectively two-dimensional) ferromagnetic phase.
 
In order to study the evolution of the quantum field during the
quench, we employ a number-conserving mean-field ansatz
\cite{mean-field} 
$\hat\psi_z=(\psi_{\rm co}+\delta\hat\psi_z)\hat A/\sqrt{\hat N}$,
which is adapted to the initial paramagnetic phase but can be
extrapolated for some finite time after the quench \cite{BH},
as long as the quantum fluctuations are small enough
$\delta\hat \psi_x, \delta\hat\psi_y, \delta\hat\psi_z \ll \psi_{\rm co}$.
This ansatz allows for a linearized expression for the magnetization
(which assumes in Cartesian coordinates the simple form
$\hat{F}_a=-i\varepsilon_{abc}\hat{\psi}_b^\dagger\hat{\psi}_c/\varrho$)
and to derive an effective mean-field Hamiltonian for transverse
magnetization $\hat{\bm F}$ 
from the full Hamiltonian Eq.\,(\ref{Hamilton}), 
\begin{equation}
\frac{\hat{\cal H}_{\rm eff}}{\varrho}
=
\frac{\hat{\bm\Pi}}{\varrho}
\left(q-\frac{\na^2}{2m}\right)
\frac{\hat{\bm\Pi}}{\varrho}
+
\hat{\bm F}
\left( \frac{q + 2c_2\varrho}{4} - \frac{\na^2}{8m} \right)
\hat{\bm F}
\,,
\label{Heff}
\end{equation}
with 
$\hat{\bm\Pi}=(\psi_{\rm co}\hat\psi_y^\dagger+\psi_{\rm co}^*\hat\psi_y\,,
-\psi_{\rm co}\hat\psi_x^\dagger-\psi_{\rm co}^*\hat\psi_x )/2$ 
being the canonical momentum operator for spin-one bosons. 
The experiment \cite{Sadler} is described by a two-dimensional
transverse magnetization $\hat{\bm F}=(\hat{F}_x,\hat{F}_y)$ obeying
the $SO(2)$-symmetry of rotations around the $z$-axis in effectively
two spatial dimensions [$\na=(\partial_x,\partial_y)$], permitting
topological defects in the form of spin vortices.
However, the above expressions for the Hamiltonian
are completely analogous for a
$SO(N)$-symmetry of the order parameter $\langle\hat{\f{F}}\rangle$ in
$N>2$ spatial dimensions, where the analogous topological defects are
generically called   ``hedgehogs'', see, e.g.,
\cite{GrishaUniverse,Mermin,Abanov,Savage}.   
Therefore, we shall discuss the general $SO(N)$ situation in the
following and return to the experimentally realized
example $N=2$ of \cite{Sadler} at the end.
 
\begin{figure}[!tb]
\includegraphics[angle=0,width=0.4\textwidth]{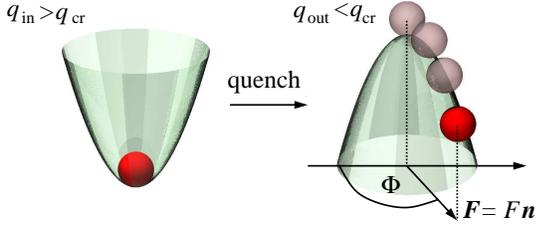}
\caption{Inversion behavior of the effective potential
from~\eqref{Heff}. 
After crossing the critical point, the direction $\bm n$ of the
effective magnetization $\bm F$ in the plane, at any given point in
space, gets frozen at a particular angle $\Phi$, while its modulus
$F$ continues to grow exponentially in time
(``rolls down'' the potential hill). 
\label{potential}}
\vspace*{-1em}
\end{figure}

The simple expression (\ref{Heff}) allows us to directly read off the
critical value $q_{\rm cr}=2|c_2|\varrho$ of the phase transition.
After a quench from $q_{\rm in}>q_{\rm cr}$ to
$q_{\rm out}<q_{\rm cr}$, and using a normal-mode expansion,
we obtain exponentially growing modes corresponding to imaginary frequencies
$\omega_k^q=\sqrt{(\epsilon_k + q) (\epsilon_k + q + 2c_2\varrho_0)}$,
where $\epsilon_k=k^2/(2m)$, cf.~Fig.~\ref{potential}.
There are two regimes:
For $q_{\rm cr}>q_{\rm out}>q_{\rm cr}/2$, the exponential growth rate
$\Im(\omega_k^q)$ of the normal modes increases with their wavelength,
but for $q_{\rm out}<q_{\rm cr}/2$, the growth rate assumes its
maximum at a given wavenumber
$k_{\rm max} = \sqrt{m( q_{\rm cr} - 2 q_{\rm out})}$.
In view of the experimental parameters \cite{Sadler},
we shall focus on the second case.
Note that due to the emergence of this preferred length scale, which
is independent of the quench time, the present situation is somewhat
different from the ``standard'' Kibble-Zurek scenario
\cite{KZ,Hindmarsh}, which is based on the
interplay between quench time, relaxation rate, and correlation
length. 

Assuming the Bose gas to be sufficiently dilute, we may extrapolate
our linearized (mean-field) description (\ref{Heff}) to intermediate
times $t$ which contain many $e$-foldings
$\Im(\omega_{k_{\rm max}}^q)t\gg1$ of the exponentially growing modes 
\cite{BH}, whereas the magnetization is still small enough, 
$\bm F^2\ll1$, to stay in the linear regime.
After a normal-mode expansion, the expectation value
$\langle\hat F_a(\bm r) F_b(\bm r') \rangle$ contains exponentially
growing contributions $\exp\{\pm i \omega_k^q t\}$.
In the limit $\Im(\omega_{k_{\rm max}}^q)t\gg1$, the $d^Nk$-integrals
are dominated by the fastest-growing modes and can thus be approximated
via the saddle-point method.
Since the initial state (paramagnetic phase) is supposed to be globally 
isotropic such as the thermal ensemble or the ground state of
(\ref{Heff}), the dominant contribution to the two-point magnetization 
correlation function is calculated to be
\bea
\label{two-point}
\langle\hat F_a(\bm r)\hat F_b(\bm r')\rangle
=
C_{\nu}^{\rm max}
\delta_{ab}
\frac{\exp\{q_{\rm cr}t\}}{\sqrt{t}}
\frac{J_{\nu}(k_{\rm max} |\bm r - \bm r'|)}
{(k_{\rm max}|\bm r - \bm r'|)^\nu}
\,,
\ea
where $J_\nu$ is the Bessel function of the first kind with the index
$\nu=N/2-1$.
(The two-dimensional case was also considered in \cite{Lamacraft,Sadler}.)
Note that the above result is universal: Apart from $\nu$, 
$q_{\rm cr}$ and $k_{\rm max}$, all information enters the pre-factor
$C_{\nu}^{\rm max}$ only, which depends on $q_{\rm out}$, sweep rate,
and temperature etc.

Now let us study the creation of topological defects, i.e.,
$SO(N)$-hedgehogs, by the symmetry-breaking quench.
Their characteristic size, i.e., the extent of the
paramagnetic core, can be determined by comparing the kinetic
term $\propto \na^2$ with the $c_2$-contribution in (\ref{Hamilton})
which yields $\xi_s=1/{\sqrt{2|c_2|\varrho m}}=1/{\sqrt{m q_{\rm cr}}}$,
i.e., the spin healing length.
In order to deal with  well-separated and therefore well-defined
topological  defects, we assume that the dominant wave-length
$\lambda_{\rm max}=2\pi/k_{\rm max}$
(determining their typical distance, see below) is much larger
than the spin healing length $\xi_s$, which
amounts to requiring $0<q_{\rm cr}-2q_{\rm out}\ll q_{\rm cr}$.
This condition is reasonably well satisfied in the experiment
\cite{Sadler} where $\xi_s\approx2.4\,\mu\rm m$ and
$\lambda_{\rm max}\approx16\,\mu\rm m$.
Within the saddle-point approximation (i.e., focusing on
the dominant modes), the term $-\na^2\to k_{\rm max}^2$ in the
equations of motion resulting from (\ref{Heff}) can be neglected
compared to $q_{\rm cr}\gg k_{\rm max}^2/m$, and we may approximate
$\ddot{\bm F}\approx q_{\rm cr}^2\bm F/4$.
Consequently, at each spatial position, the magnetization $\bm F$
behaves as the coordinate of a harmonic oscillator
(in $N$ dimensions), becoming inverted upon crossing the transition,
cf.~Fig.~\ref{potential}.
Splitting $\bm F$ up into its modulus $F$ and unit vector of
direction $\bm n$ via $\bm F=F\bm n$, we find that $F$ grows
exponentially whereas the dynamics of $\bm n$ freezes:
In view of the conserved ``angular momentum''
$\bm L={\bm F}\times\dot{\bm F}$, we find that
$|\dot{\bm n}|=|\bm L|/F^2\propto F^2_0/F^2(t)$ decreases rapidly.
Ergo, after a short time (of order $1/q_{\rm cr}$), the magnetization
$\bm F$ at each spatial position $\bm r$ grows exponentially in some
given direction $\bm n(\bm r)$, i.e., the system rolls down the
parabolic potential hill whereby the direction of descent does not
change anymore, cf.~Fig.~\ref{potential}.
The correlations between the frozen directions $\bm n$ at different
spatial positions are governed by the initial state and determine the
seeds for the creation of topological defects -- i.e., objects which
cannot be deformed to a constant $\bm n$ in a smooth way
\cite{Mermin}. 
 
The topological defects we are considering are $SO(N)$
(anti-)hedgehogs in $N$ spatial dimensions with the typical order
parameter distribution $\bm n=\pm\bm r/r$ and
can be characterized by a non-vanishing winding number,
i.e., a topological invariant or ``charge''
belonging  to the homotopy group
$\pi_{N-1}(S_{N-1})= {\mathbb Z}$.  
The winding number is calculated by an integral over a
$N-1$-dimensional hypersurface enclosing a $N$-dimensional volume
\cite{Abanov}: 
\bea
\label{winding}
\hat{\mathfrak N}
=
\oint dS_\alpha\,
\frac{\varepsilon_{abc\dots}\varepsilon^{\alpha\beta\gamma\dots}}
{\Gamma(N) \,{\cal S}_{N-1}}\,
\hat n^a(\partial_\beta\hat n^b)(\partial_\gamma\hat n^c)\dots
\,.
\ea
Here $\varepsilon$ denote the Levi-Civita symbols and 
${\cal S}_{N-1}= 2\pi^{N/2}/\Gamma(N/2)$
the surface of the unit sphere 
in $N$ dimensions.
Since the winding number operator 
counts the difference between hedgehogs and
anti-hedgehogs, its expectation value vanishes, 
$\langle\hat{\mathfrak N}\rangle=0$, 
but its variance $\langle\hat{\mathfrak N}^2\rangle$ as a function of
the enclosed volume yields the desired spectrum of net magnetization
fluctuations.
 
In order to calculate $\langle\hat{\mathfrak N}^2\rangle$, we make
an additional approximation:
For a $N$-dimensional harmonic oscillator
$\ddot{\bm F}\approx q_{\rm cr}^2\bm F/4$, the ground-state
probability distribution $p(F)\propto F^{N-1}\exp\{-F^2\}$
of the amplitude $F$ is peaked around a (for $N>1$) non-zero value
and the relative width of this peak decreases with increasing $N$. 
Therefore, we approximate the operator $\hat F$ by an exponentially
growing classical value $\hat F\to F(t)$ in all expectation values.
This semiclassical approximation will become asymptotically exact in
the limit $N\uparrow\infty$, but we expect it to yield qualitatively
correct results also in lower dimensions down to $N=2$, since $p(F)$ is
discernibly peaked even for $N=2$ and the topological defects are
mainly determined by the (quantum) fluctuations of $\hat {\bm n}$
(and not of $\hat F$).
For example, calculating [analogously to (\ref{two-point})] 
the contribution of the fastest growing modes to the correlator 
$\langle\hat{\bm F}^2(t,\bm r)\hat{\bm F}^2(t,\bm r')\rangle=F^4(t)+
2\langle\hat{\bm F}(t,\bm r)\cdot\hat{\bm F}(t,\bm r')\rangle^2/N$,
we see that the accuracy of the semiclassical approximation 
$\hat F\to F(t)$ increases for large distances $|\bm r-\bm r'|$ and/or
large $N$.
Note that all the linearized $\hat F_a$ possess independent initial
ground states and commute with each other.
Therefore, the $\hat n_a$ do also commute with each other [so that
operator ordering in \eqref{winding} is not an issue] and with $\hat F$, 
which ultimately makes possible the semiclassical approximation of
$\hat F \rightarrow F$.
 
\begin{figure}[!tb]
\includegraphics[angle=0,width=0.4\textwidth]{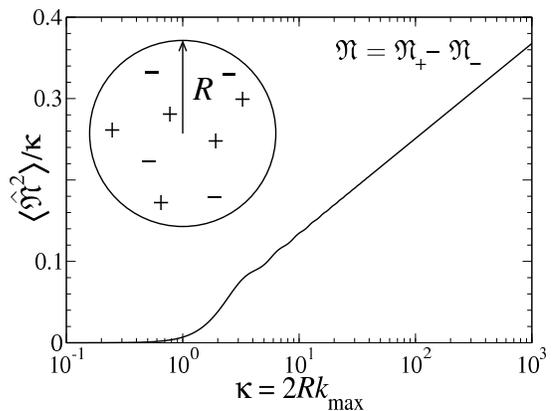}
\caption{Scaling of the variance of the winding number
$\langle\hat{\mathfrak N}^2\rangle$ with the sample size $R$. 
A logarithmic plot of $\langle\hat{\mathfrak N}^2\rangle/\kappa$ 
is used in order to visualize the asymptotic $\kappa\ln\kappa$ scaling. 
\label{N2}}
\end{figure}

The above approach enables us to calculate
$\langle\hat{\mathfrak N}^2\rangle$ in arbitrary dimensions $N\geq2$.
After inserting (\ref{winding}) and using the semiclassical
approximation $\hat{\bm F}(t,\bm r)\approx F(t)\,\hat{\bm n}(\bm r)$,
the expectation value can be decomposed into a product of $N$
two-point correlators (\ref{two-point}).
Let us exemplify this procedure for two spatial dimensions:
The transverse magnetization can be decomposed via
$\hat F_\perp = \hat F_x +i \hat F_y = \hat F e^{i \hat \Phi}$
into its modulus $\hat F$ and phase $\hat \Phi$ which commute and are
both self-adjoint \cite{remark}.
The winding number (\ref{winding}) reads \cite{remark2}
\bea
\hat{\mathfrak N}
=
\frac{1}{2\pi}
\oint \limits_{\mathfrak C} d\bm l \cdot \na \hat \Phi
=
\oint \limits_{\mathfrak C} d\bm l \cdot
\frac{\hat F_x \na \hat F_y - \hat F_y \na \hat F_x}{2\pi\hat F^2}
\,.
\label{wind}
\ea
Using Eq.\,\eqref{two-point}, replacing $\hat F \rightarrow F$,
and choosing the contour $\mathfrak C$ as a circle with radius $R$,
we get \cite{small}
\bea
\langle\hat{\mathfrak N}^2(R) \rangle
=
\frac{\kappa^2}{2\pi}
\int\limits_0^1 dx\,\sqrt{1-x^2}\,J_1^2(\kappa x)
\,, \label{J1}
\ea
with $\kappa = 2 R k_{\rm max} $, cf.~Fig.\,\ref{N2}.
As anticipated, the characteristic length scale
(e.g., typical distance between vortices) is set by the dominant 
wavelength $\lambda_{\rm max}$
(instead of the spin healing length $\xi_s$, for example).
 
The analytical expression (\ref{J1}) enables us to study the large-scale
spectrum of the winding number, i.e., the difference between the
number of vortices and anti-vortices 
${\mathfrak N}={\mathfrak N}_+-{\mathfrak N}_-$.
Two models for the scaling behavior of ${\mathfrak N}$ 
are commonly discussed in the literature \cite{Hindmarsh,GrishaUniverse}. 
Assuming a random distribution of vortices and anti-vortices
({\em random vortex gas} model), the typical difference 
$|{\mathfrak N}|$ scales with $\sqrt{{\mathfrak N}_\pm}$.
Because the total number ${\mathfrak N}_++{\mathfrak N}_-$ increases 
proportional to the area $R^2$, this model predicts a scaling of
$\langle\hat{\mathfrak N}^2\rangle\sim R^2$ ($N=2$).
Alternatively, the {\em random phase walk} model assumes a random walk 
of the phase $\Phi$ along the circumference.
The accumulated winding number $|{\mathfrak N}|$ scales
with $\sqrt{R}$ in this situation, implying a
scaling of $\langle\hat{\mathfrak N}^2\rangle\sim R$.
Of course, both models cannot be the final truth (consider the
deformation of the circle to a slim ellipse, for example), and it is
not obvious how to generalize the random phase walk model to higher
dimensions -- but one can compare the different predictions with our
approach.
For large $\kappa$, the expansion
$J_1(\kappa x\uparrow\infty)\propto\cos(\kappa x-3\pi/4)/\sqrt{\kappa x}$
yields (cf.~Fig.\,\ref{N2})
\bea 
\label{log-corr}
\langle\hat{\mathfrak N}^2(\kappa \gg 1)\rangle 
\sim \kappa\ln\kappa
\,,
\ea
which is close to the scaling predicted by the random phase walk model 
-- but still not quite in agreement due to the additional factor of 
$\ln\kappa$. 
This extra factor only occurs for $N=2$ and arises from the slow decay
of the two-point correlator (\ref{two-point}) at large distances 
$|\f{r}-\f{r}'|^{-1/2}$, which deviates from a local random phase
walk \cite{remark3}. 
Although the sample size in \cite{Sadler} is probably not sufficient
for extracting the factor $\ln\kappa$, this prediction 
could be tested in future experiments. 

Let us study intermediate $\kappa$ values and compare our
results with the experiment \cite{Sadler} by placing a circle with a
radius of $R=5\,\mu\rm m$ into the center of the condensate cloud.  
Since $R$ is significantly bigger than the spin healing length 
$\xi_s\approx2.4\,\mu\rm m$ but still sufficiently below the planar
Thomas-Fermi radii of the cloud 
($12.8\,\mu\rm m$ and $167\,\mu\rm m$), the assumptions entering our   
derivation (such as homogeneity) should provide a reasonably good
approximation. 
For $R=5\,\mu\rm m$ corresponding to $\kappa\approx4$, we obtain  
$\langle\hat{\mathfrak N}^2\rangle\approx1/3$, i.e., the probability 
of finding an (anti-)vortex inside the circle is around $1/3$. 
In view of the cigar shape of the cloud, we may place several circles
in the central region of the condensate such that their mutual
distance exceeds the correlation length (typical domain size) of  
$\lambda_{\rm max}/2\approx8\,\mu\rm m$. 
Since these circles can be considered nearly independent, we would
expect at least one vortex in the cloud -- in reasonable agreement
with \cite{Sadler}. 
In contrast, the defect density is calculated in Ref.~\cite{Lamacraft} 
to be $k_{\rm max}^2/(4\pi)$, treating the magnetization as a Gaussian
stochastic field. 
Therefore, a circle with a radius of $R=2/k_{\rm max}\approx5\,\mu\rm m$
should typically contain one defect.
Although the discrepancy ($1/3$ vs.\ one) is not huge, it already
suggests that our results are not quite compatible with
Ref.~\cite{Lamacraft}, and thus experiment should judge which approach
yields the better description.
Alternatively, numerical simulations (supplemented by suitable
assumptions about the initial noise spectrum) provide a complementary
approach \cite{SaitoBerkeley}.

In conclusion, for the example of a spinor Bose gas, we studied the
creation of topological defects during the quench from the
paramagnetic to the $SO(N)$-symmetry breaking ferromagnetic phase.
For an arbitrary number $N$ of spatial dimensions, we derived a
universal expression for the winding number (which only depends on $N$
and the enclosed volume in units of 
 $\lambda_{\rm max}^N$) and found a
logarithmic correction to the scaling law of the random phase walk
model for $N=2$.  

R.\,S. and M.\,U. acknowledge support by the Emmy Noether Programme of
the German Research Foundation (DFG) under grant No.~SCHU~1557/1-2.
We thank D.\,M.~Stamper-Kurn and G.\,E. Volovik for valuable
discussions, and acknowledge support by ESF-COSLAB. 
\\ 
$^*$\,{\footnotesize\sf schuetz@theory.phy.tu-dresden.de}\,\,   
$^\dagger$\,{\footnotesize \sf uwe.fischer@uni-tuebingen.de}



\begin{thebibliography}{9999}

\newfont{\cyr}{wncyr10}

\bibitem{KZ} 
T.\,W.\,B.~Kibble, J.~Phys A {\bf 9}, 1387 (1976);
W.\,H.~Zurek, Nature {\bf 317}, 505 (1985).~

\bibitem{Hindmarsh}  
M.~Hindmarsh and T.\,W.\,B.~Kibble,
Rep.~Prog.~Phys.~{\bf 58}, 477 (1995);  
W.\,H.~Zurek, Phys.~Rep.~{\bf 276}, 177 (1996).

\bibitem{ZDZ} 
W.\,H. Zurek, U. Dorner, and P. Zoller, 
Phys. Rev. Lett. {\bf 95}, 105701 (2005).

\bibitem{Stenger} 
J.~Stenger {\it et al.}, 
Nature {\bf 396}, 345 (1998).~

\bibitem{M-S_Chang} 
M.-S.~Chang {\it et al.}
Nature Phys.~{\bf 1}, 111 (2005).~

\bibitem{Zhang} 
W.~Zhang {\it et al.}, 
Phys.~Rev.~Lett.~{\bf 95}, 180403 (2005).

\bibitem{Sadler} 
L.\,E.~Sadler, J.\,M.~Higbie, S.\,R.~Leslie,
M.~Vengalattore,  and D.\,M.~Stamper-Kurn,
Nature {\bf 443}, 312 (2006).

\bibitem{polar} 
For brevity, we shall use the terms para- and ferromagnetic
synonymously for the more precise descriptions polar paramagnetic and
planar ferromagnetic, respectively.  

\bibitem{Higbie} 
J.\,M.~Higbie {\it et al.},
Phys.~Rev.~Lett.~{\bf 95}, 050401 (2005).

\bibitem{Saito} 
H.~Saito and M.~Ueda, Phys.~Rev.~A {\bf 72}, 023610 (2005).

\bibitem{Chiral} 
H. Saito, Y. Kawaguchi, and M. Ueda,
Phys. Rev. Lett. {\bf 96}, 065302 (2006).
 
\bibitem{Ohmi} 
T.~Ohmi and K.~Machida,
J.~Phys.~Soc.~Jpn.~{\bf 67}, 1822 (1998); 
T.-L.~Ho,
Phys.~Rev.~Lett.~{\bf 81}, 742 (1998).

\bibitem{mean-field}
Here $\hat N=\hat A^\dagger\hat A$ counts the total number of
particles, cf.~C.\,W.~Gardiner,
Phys.\ Rev.\ A {\bf 56}, 1414 (1997);
%
Y.~Castin and R.~Dum,
Phys.\ Rev.\ A {\bf 57}, 3008 (1998).

\bibitem{BH} 
R.~Sch\"utzhold, M.~Uhlmann, Y.~Xu, and U.\,R.~Fischer,
Phys.~Rev.~Lett.~{\bf 97}, 200601  (2006).

\bibitem{GrishaUniverse} 
G.\,E.~Volovik and V.\,P.~Mineev,
Zh.~\' Eksp.~Teor.~Fiz.~{\bf 73}, 767  (1977) 
[Sov.~Phys.~JETP {\bf 46}, 401 (1977) contains a misprint in Eq.\,(2.4); 
in the denominator ``6'', i.e., $\Gamma(4)$, was replaced by ``8'']; 
G.\,E.~Volovik, {\em The Universe in a Helium Droplet}  
(Oxford University Press, 2003).

\bibitem{Mermin} 
N.\,D.~Mermin,
Rev.~Mod.~Phys.~{\bf 51}, 591-648 (1979); L.~Michel,
Rev.~Mod.~Phys.~{\bf 52}, 617-651 (1979).~

\bibitem{Abanov} 
A.\,G.~Abanov and P.\,B.~Wiegmann,
Nucl.~Phys.~B {\bf 570}, 685 (2000).

\bibitem{Savage} C.\,M.~Savage and J.~Ruostekoski,
Phys.~Rev.~A {\bf 68}, 043604 (2003).~

\bibitem{Lamacraft} A.~Lamacraft,
Phys. Rev. Lett. {\bf 98}, 160404 (2007). 

\bibitem{remark}
Note that this is distinct from the nonvanishing commutator of density
and phase operators resulting from the analogous decomposition of the
single-component field operator 
$\hat\psi = \exp [i\hat\Phi ]\sqrt{\hat\varrho}$, which underlies
the definition of conventional momentum/velocity vortices in a fluid.

\bibitem{remark2}
In contrast to spin vortices, domain walls (lines with vanishing $F$)
are not topologically stable for the order parameter considered here.
%
Thus, the domain boundaries observed in \cite{Sadler} are line-like
regions with small but not vanishing $F$, and do not enter
Eqs.~\eqref{winding} and \eqref{wind}. 

\bibitem{small}
The limit of very small $\kappa$ requires further consideration, 
which will be presented elsewhere. 

\bibitem{remark3}
Observe that the extra factor of $\ln\kappa$ in Eq.~(\ref{log-corr}) 
is only valid within the saddle-point approximation (\ref{two-point}), 
for large times $t$ (but still in the linear regime
\cite{initial}). 
%
Conversely, 
keeping the (large) time $t$ fixed, and considering the limit 
of ultra-large distances $R$, 
the scaling levels off to $\kappa$. 

\bibitem{SaitoBerkeley} 
H.~Saito, Y.~Kawaguchi, and M.~Ueda,
Phys. Rev. A {\bf 75}, 013621 (2007); 
%
{\sf preprint} arxiv:0704.1377 [cond-mat.other]. 

\bibitem{initial}
Note that our results do only describe the initial stage and are,
strictly speaking, only valid in the linear regime, i.e., as long as the
directions $\bm n(\bm r)$ are frozen.
%
The subsequent non-linear dynamics, including annihilations of
vortex--anti-vortex pairs etc., is not taken into account. 

\end{thebibliography}
\end{document}